\documentclass[preprint2]{aastex}

\slugcomment{manuscript for AJ, vers.3, 100322}

\shorttitle{UCAC3 pixel reduction paper}
\shortauthors{Zacharias}

\begin{document}

\title{UCAC3 pixel processing}



\author{Norbert Zacharias$^1$}

\email{nz@usno.navy.mil}

\affil{$^1$U.S.~Naval Observatory, 3450 Mass.Ave.~NW, Washington DC 20392} 


\begin{abstract}
The third US Naval Observatory (USNO) CCD Astrograph Catalog, UCAC3
was released at the IAU General Assembly on 2009 August 10.
It is a highly accurate, all-sky astrometric catalog of about 100 million
stars in the R = 8 to 16 magnitude range.  Recent epoch observations are 
based on over 270,000 CCD exposures, which have been re-processed for the
UCAC3 release applying traditional and new techniques.
Challenges in the data have been high dark current and asymmetric image
profiles due to the poor charge transfer efficiency of the detector.
Non-Gaussian image profile functions were explored and correlations
are found for profile fit parameters with properties of the CCD frames. 
These were utilized to constrain the image profile fit models and 
adequately describe the observed point-spread function of stellar 
images with a minimum number of free parameters.
Using an appropriate model function, blended images of double
stars could be fit successfully.
UCAC3 positions are derived from 2-dimensional image profile fits
with a 5-parameter, symmetric Lorentz profile model. 
Internal precisions of about 5 mas per coordinate and single
exposure are found, which are degraded by the atmosphere to
about 10 mas.  However, systematic errors exceeding 100 mas
are present in the $x,y$ data which 
have been corrected in the astrometric reductions following
the $x,y$ data reduction step described here.
\end{abstract}


\section{INTRODUCTION}

The US Naval Observatory (USNO) operated the 8-inch (0.2 m) Twin
Astrograph between 1998 and 2004 for the first ever all-sky astrometric 
survey using a CCD detector.
About 2/3 of the sky was observed from the Cerro Tololo Inter-American
Observatory (CTIO) while the rest of the northern sky was observed
from the Naval Observatory Flagstaff Station (NOFS).
In 1998 the Kodak 4k by 4k CCD detector used for this UCAC project 
was the largest number of pixels chip at any telescopes at CTIO.
The UCAC project also brought attention to a potential source of
systematic errors for astrometry using CCDs, the charge-transfer
inefficiency.

The first paper describing the UCAC1 release \citep{ucac1} gives 
details about the observing procedures and initial reductions.  
The second release, UCAC2, \citep{ucac2}, is an extension of UCAC1 
applying similar reduction methods to a much larger area of
the sky. The same pixel processing pipeline was used for UCAC1
and UCAC2, while improved systematic error corrections were
introduced for the UCAC2 reductions to obtain celestial
coordinates.
The UCAC2 is being used extensively by the astronomical community,
providing a much needed densification of the optical celestial
reference frame at magnitudes fainter than the Hipparcos and Tycho-2 
catalogs.
The UCAC observations cover mainly the 8 to 16 magnitude range,
providing accurate star positions with external errors of about
20 to 70 milliarcsecond (mas) depending on magnitude.

For the recent UCAC3 release \citep{ucac3} completely new reductions
of the pixel data were performed, involving new analysis methods.
These reductions and results are described in this paper in detail,
including image profile fitting methods leading to $x,y$ centers.
The subsequent astrometric processing from $x,y$ data to celestial
coordinates is described in a separate paper \citep{ared}.

Point-spread function (PSF) fitting has been performed in the past,
see for example \citep{hstpsf} for HST data, or the IRAF DAOPHOT package
\citep{irafpsf}.  The approach taken here is different, deriving
relatively simple, analytical model functions which describe the
observed PSF sufficiently well. At the same time the number of free 
parameters needed for each image profile fit is kept to a minimum by
utilizing information from many CCD exposures to constrain some
image profile model parameters.
Challenges here are asymmetric PSFs and variations of the PSFs over
the field of view, combined with a relatively small number of stars
per CCD frame and the goal of high astrometric accuracy.

\section{PIXEL DATA}

A 4094 by 4094 pixel CCD with a 9 $\mu$m pixel size was used in a single
bandpass (579 to 643 nm) providing a field of view (FOV) of just over
1 square degree, taking advantage of only a tiny fraction of the flat FOV
delivered by the optical system of the Twin Astrograph's ``red lens.''
This camera provides 14 bit output and has a gain setting of 
5.65 electrons (e$^{-}$) per analog-to-digital unit (ADU), 13 e$^{-}$ 
read noise, and about 85,000 e$^{-}$ full well capacity.

A 2-fold overlap pattern of 85,158 fields spans the entire sky.
Each field was observed with a long (about 125 s) and a short
(about 25 s) exposure.
The raw pixel data are stored in custom FITS differential compress
(fdc) format files, about 16 MB per exposure without loss of the 
14 bit dynamic range.

The detector features a poor charge transfer efficiency (CTE)
leading to asymmetric images along the readout direction 
($x$ axis) which vary as a function of distance from the output register.
This leads to systematic errors in the star positions as a function
of $x$ and the stars' brightness (magnitude), 
about the worst thing what can happen for an astrometric instrument.  
The contour plots in Fig.~1 illustrate this problem showing the
change of image shape (from almost circular to pronounced asymmetric) 
as seen on the left and right side of the detector, respectively.
The left side (low $x$) is close to the readout register and also
displays the largest background noise on the chip, likely due to
a higher than average temperature there. 
Initially the camera showed a ``glowing spot" in the lower left
corner.  The design was changed to have the read amplifier powered 
up only when needed, which eliminated that problem.

In order to mitigate these $x$ and magnitude dependent systematic 
errors the detector was operated at a relatively high temperature 
($-18 C$), which filled many of the CTE causing traps on the 
silicon detector.  Unfortunately the warm operating temperature 
leads to a substantial dark current.
Frequent darks were taken throughout the project for each of the
standard exposure times (5, 10, 20, 25, 30, 40, 100, 125, 150, 200 s).
Some time into the project it was discovered that the darks also
depend on ambient temperature and vacuum pressure inside the camera,
which due to small leaks increased from about 0.1 torr to over 2 torr,
when a new pumpout of the camera was performed every few months.

\section{RAW DATA PROCESSING STEPS}

\subsection{Combined darks and bad pixel map}

The detector used for the UCAC survey has a high cosmetic quality
with no bad columns and relatively few bad pixels.
In order to simplify the reductions and assuming the worst case, 
a single list of all possible bad pixels were established spanning 
dark exposures taken during the entire project. 

Early on it was discovered that darks taken during daytime or in rapid
succession display different properties than object frames taken 
during regular observing.  
Most darks therefore were taken during cloudy nights with a script
to obtain about 50 darks of a given integration time in an automated 
sequence.  Pauses of about a minute between dark exposures were 
introduced to closely resemble actual observing conditions.
Using custom software these 50 FITS files were read in parallel, block by
block and the 50 measures of each pixel sorted.  The mean pixel value was
calculated after rejecting about 10 \% of the lowest and highest values.
This way every few weeks a new combined dark was constructed for every
standard exposure time used during that period.

To identify bad pixels comprehensive samples of combined darks of
a given exposure time were compared, pixel by pixel.  If either the 
scatter or the mean pixel value exceeded adopted thresholds (about 
3-sigma level), that pixel was flagged as ``bad".
This process was repeated for all exposure times and a combined
list of pixels generated of those pixels which appeared at least 
once on any of the individual ``bad" pixel lists.
A total of 13,094 such pixels was identified, which is less than
0.1 \% of all pixels on the detector.

\subsection{Applying darks}

The average background intensity (from bias and dark current) of
raw CCD frames taken with our 4k camera is very nonuniform over
the field.  However, the pattern is very similar from exposure to
exposure, while the amplitude of the pattern depends on many things,
like exposure time, ambient temperature and vacuum pressure.
The mean difference in background ADU between the left and right
side of the CCD frame serves as a parameter to quantify the
amplitude of this background pattern.

The re-processing of the pixel data was split up into batches of
about 10,000 consecutive CCD frames taken over a narrow range of 
epochs.  A pair of appropriate master darks for each standard exposure 
time was selected.  Each pair spans a range in background differences
(left to right, see above) that is as large as possible with the 
restriction of being taken close to the epoch of the frames under 
investigation.
The raw data processing then involved a determination of the mean
background difference (left to right) of each individual frame.
This value was used in a linear interpolation between the 2 master 
darks selected for that set of data and the exposure time of the
object frame.  The pixel-by-pixel interpolated dark was then
subtracted from the object frame.

This method of dark subtraction was new for UCAC3 and resulted in
a significant improvement in background flatness and lower noise,
which leads to a deeper and more uniform limiting magnitude than 
before.  For UCAC2 a more or less random pick of a dark near the 
target properties was selected without interpolation.
As with previous releases, no additional bias frames were needed.

\subsection{Flats}

Due to the small size of the field utilized by the CCD, as compared
to the optical design of the astrograph, there is no vignetting from
the optical system expected. Initial tests also revealed only small 
pixel-to-pixel sensitivity variations.  The window on the camera 
serves as the only filter in a sealed system without moving parts.
Thus for the UCAC1 and UCAC2 releases no flats were applied
at all to the survey data, aiming at astrometric results without
the goal of precise photometry.

However, a set of about 25 dome flats were taken every few months
with an exposure time of 5 s and light intensity set to give
about 30 to 50\% full well capacity illumination.
These data were reduced and applied for the UCAC3 release.
The appropriate combined dark frame was subtracted from each individual
flat exposure, and all flats of a given epoch were combined excluding
extreme low and high counts, similarly to the darks processing
described above.
The flats were scaled to 1000 ADU mean intensity (integer)
representing a factor of 1.0 for the science frames data processing 
to follow.
For some epochs the flat data were split into 2 groups of high/low
average illumination to check on internal consistency.  A total of 
28 combined master flats was thus obtained spanning the entire 
duration of the UCAC observing.

The pixel-to-pixel sensitivity variations are small. 
Taking 1/25 of the entire CCD area at a time, sorting all the pixels
of a given master flat and cutting the low 5\% and high 5\% of the pixels,
the resulting standard deviation for pixel-to-pixel variation is only on
the order of 0.4 to 0.6\% of the mean pixel count.
This fact explains why excellent astrometric results (center fit precision
close to 1/100 pixel) were obtained in UCAC2 even without applying any 
flats.

Large-scale sensitivity variations over the CCD frame area were found 
to be 10\% or less.
Comparing different master flats of different epochs, variations
of about 2\% or less are found, except for the set taken around night
numbers 2000 to 2150 (truncated Julian Dates), where significant 
deviations due to a shutter failure problem are found in the corners
of CCD frames.

Based on these results a single master flat file close to the
epoch of each object frame was selected and applied.
The $\approx$ 10\% vignetting in the corners of the CCD frames was
attributed to a slightly undersized, round opening of the shutter 
in our 4k camera system.

\section{IMAGE PROFILES}

\subsection{Supersampling}

In order to investigate the shape of the point-spread function (PSF)
as seen in the UCAC data the following standard procedure was adopted.
Individual CCD frames of good quality taken in areas of the sky with
large numbers of stars (but not too crowded, few blended images) were
selected.
Centers of stellar images were determined by least-square fits using
a 2-dimensional Gaussian model of the pixel intensity ($I$) as a
function of the pixel coordinates ($x, y$),

\begin{equation}
I (x,y) = B + A \ e^{-\frac{r^{2} ln(2)}{r_{0}^{2}}}
\end{equation}

\noindent with $r^{2} = (x - x_{0})^{2} + (y - y_{0})^{2}$.
We call this model 1 and it has 5 free parameters: the average local 
background intensity ($B$), the amplitude ($A$), the width of the 
profile ($r_{0}$), and the image center coordinates $x_{0}$ and $y_{0}$.
The natural logarithm of 2 is included here to scale the $r_{0}$ to
obtain the radius of the profile at half maximum.

Images with sufficient signal-to-noise (S/N) ratio, but not saturated,
were scaled to a fixed amplitude, shifted in $x,y$ pixel coordinates
to align centers and resampled onto a grid with 0.2 pixel resolution.
Averages of the pixel values in each bin were taken to produce the
supersampled PSF representative for that particular CCD frame or
sub-area of it.
For some of the investigations presented below, a CCD frame was
split into 3 equal area sections along the $x$-axis, following
increased effects of the poor CTE of the detector.
Marginal profiles were calculated from these 2-dim supersampled
PSFs along the $x$ and $y$ axis, labeled as $u$ and $v$ coordinates.
A 1-dim radial profile was generated for 0.2 pixel bins using the
original pixel data and assuming a circular symmetric intensity
distribution.  The spatial coordinate for those profiles is called 
$r$, as defined above.

\subsection{Profile model functions}

An overview of all profile models used for UCAC3 reductions
and tests is given in Table 1.
Models 1 through 4 are taken from the software for analyzing 
astrometric CCD data (SAAC) \citep{winter}, which developed
out of earlier work \citep{schramm},  while the other
models have been modified from SAAC models or are newly developed.
Model 1 is given by Eq.~1, while model 2 did perform better than
model 1 but worse than model 4 and is not further discussed here.
Model 4 is the general Lorentz profile function, given as

\begin{equation}
I (x,y) = B \ + \ A \ \left[ 1 + \left(\frac{r}{r_{0}}\right)^\alpha \ 
                 \left(2^{1/\beta} - 1\right) \right]^{-\beta} 
\end{equation}

For $\beta = 1$ this reduces to the Moffat profile \citep{moffat}.
The additional parameter in model 4 gives more control over the
shape of the PSF, allowing adjustment of the gradient near the core
of the profile independently from the gradient out in the wings.

The profile function of model 4 with a fixed, preset parameter 
$\beta$ (not determined in the least-squares fit as a free 
parameter) we call here model 3. 
Similarly, a function with both parameters $\beta$ and $\alpha$ 
preset we call model 5.

Derived from the circular, symmetric, Gaussian profile (model 1)
an elliptical symmetric function with major and minor
axes aligned to $x$ and $y$, respectively (model 6) is defined by

\begin{equation}
I (x,y) = B + A \ e^{- ln(2) \left(\frac{(x-x_{0})^{2}}{a^{2}}
                   + \frac{(y-y_{0})^{2}}{b^{2}} \right) }
\end{equation}

\noindent where $B$, $A$, $x_{0}$, and $y_{0}$ are defined as 
before and $a$ and $b$ are the widths of the profile along the 
$x$ and $y$ axis, respectively.

Model 7 is a generalization of model 6 with the additional free 
parameter $\theta$ describing the angle of the major axis with 
the $x$ axis (range $\pm \pi/2$), and is given by

\begin{equation}
I (x,y) = B + A \ e^{- \left((x-x_{0})^{2} c_{x}
        + (y-y_{0})^{2} c_{y} + (x-x_{0}) (y-y_{0}) c_{m} \right) }
\end{equation}

\noindent with 
\begin{eqnarray}
 c_{x} =  \frac{1}{2} \left( \frac{\cos^{2}(\theta)}{a^{2}} 
      + \frac{\sin^{2}(\theta)}{b^{2}} \right) \nonumber \\
 c_{y} = \frac{1}{2} \left( \frac{\cos^{2}(\theta)}{b^{2}} 
       + \frac{\sin^{2}(\theta)}{a^{2}} \right) \nonumber \\
 c_{m} = - \cos(\theta) \sin(\theta) \left( \frac{1}{b^{2}} 
                 - \frac{1}{a^{2}} \right)  \nonumber 
\end{eqnarray}

\noindent which reduces to model 6 for $\theta$ = 0.

Tests with asymmetric profiles were performed by adding the
following term to each base model ($F$, as given above) in the 
description of the pixel intensities as function of $x,y$ 

\begin{equation}
  I (x,y) = B + A \ F(r) \ +  \ A \ c \ (x-x_{0}) \ F(r) 
\end{equation}

This adds an asymmetric part to the $x$ component with an
amplitude factor, $c$.
Extending this concept to an additional asymmetric contribution
along the $y$ axis is straightforward.

The following model was used for some tests. 
It is based on a generalized Lorentz profile (elliptical) with
asymmetric terms for $x$ and $y$,

\begin{equation}
I (x,y) = B \ + \ A^{\prime} \ \left[ 1 + \left(\frac{r}{r_{0}}\right)^\alpha \ 
                 \left(2^{1/\beta} - 1\right) \right]^{-\beta} 
\end{equation}

\noindent with
\begin{eqnarray}
  \Delta x = x - x_{0}  \nonumber \\ 
  \Delta y = y - y_{0}  \nonumber \\
  \frac{r}{r_{0}} = \sqrt{\frac{\Delta x^{2}}{a^{2}} + 
                          \frac{\Delta y^{2}}{b^{2}}}
     \nonumber \\
  A^{\prime} = A \left[ 1 + c_{x} \Delta x 
               + c_{y} \Delta y \right] \nonumber
\end{eqnarray}

\noindent with background $B$, amplitude $A$, image center
$x_{0}, y_{0}$, profile shape parameters $\alpha, \beta$ as before, 
and radius of profile width along $x$ and $y$, $a, b$, respectively 
(elliptical model).
The asymmetry in this model is described by the parameters
$c_{x}$, and $c_{y}$ for the relative amplitude
of the asymmetry along $x$ and $y$, respectively, 
giving a total of 10 parameters.
This profile function was used for models 12 to 14, depending on
which of these parameters are preset and which are free fit parameters
(see Table 1).

\subsection{First results}

Figure 2 shows the radial profile of a supersampled PSF
(as explained above) obtained from stellar images on the
left side (nearly no CTE effect) of CCD frame 173586 as an 
example.
The same data points are shown in both plots.
However, for the top plot the Gaussian (model 1) function
was used to generate the fit line through the data points,
while the bottom plot shows the result of model 4.
A much better fit to the actual data is obtained with the latter model.

The small crosses running through the middle of each plot show the
residuals (data $-$ fit model), scaled by a factor of 3 and
offset by a constant along the intensity axis for better 
visualization.
The Lorentz profile model gives significantly smaller residuals
(about a factor of 2) than the Gaussian model; however, the Lorentz
profile fit is not perfect either, and the spacial frequency of the
residuals has increased, giving more peaks and valleys in the 
residual pattern, thus increasing its complexity.

Figure 3 shows a similar set of plots obtained from the same
CCD frame; however, using stellar images on the right side of
the CCD (large CTE effect) and showing a marginal cut 
along $x$ instead of the radial coordinate used in Fig.~2.
Clearly the asymmetry of the profile is seen, and both models
perform about equally well, with a slightly better fit for the
Lorentz model.

Figure 4 illustrates contour plots of residuals after fitting the
supersampled PSF of the right side (large CTE effect) of
frame 134473 (compare to Fig.~1).
This asymmetric image could be fit reasonably well with
model 9 using an elliptical, general, Lorentz profile function and
asymmetry terms for $x$ (see Eq.~5), a total of 9 free parameters.
The contour residuals using model 4 are shown for comparison.
Residuals of a fit with model 1 look similar to the model 4 plot,
although with larger amplitudes.  In any case relatively large 
residuals with high spacial frequencies remain even when applying
the asymmetric model.

\subsection{Minimize number of free parameters}

Fitting individual stellar images which extend only over
a few pixels with models having 8 or even 10 free parameters,
if numerically feasible at all, 
will lead to poor results in astrometry due to the small
degree of overdetermination in the least-squares process.
In order to benefit from the image profile models which
better fit our data than the Gaussian some restrictions
in parameter space were investigated.

A set of 282 high quality CCD frames was selected to sample
the range in FWHM and span the entire observing epoch range.
All these frames have a large number of stars, but are not crowded.
For all frames the supersampling of the PSF was performed and
various image profile fit models run on these, separately for
each CCD frame.  Results were summarized in a table and supplemented
by observing log items.

Figure 5 shows the strongest correlation found for the various
parameters investigated.  The shape parameters ($\alpha, \beta$)
of the symmetrical Lorentz profile model 4 can be predicted from
the profile width of the Gaussian model 1 fit.
The profile width here is the radius (about FWHM/2) with unit 
bin width (0.2 pixel) of the supersampled profile data. 
A linear term is sufficient to predict the $\beta$ parameter,
while for the $\alpha$ parameter a second order polynomial 
was adopted.  Scaling to the actual pixel size these results
were hard-coded to preset both shape parameters in profile 
fit model 5, which is otherwise the same as model 4.
This leaves only 5 free fit parameters, exactly the same as for
the 2-dim Gaussian model function (see above and Table 1).

Tests were performed to determine any possible variations of
the $\alpha$ and $\beta$ parameters.
Supersampled PSFs were generated as a function of 2 magnitude
bins, and in another test the data were split into 4 quadrants
on the CCD frames.  Consistent results for the $\alpha$ and $\beta$ 
parameters were found, confirming the relationship with the
profile width as before with very small variation as a function
of other selection criteria.

\subsection{Other models}

Tests were performed using elliptical profile models (6,7,8).
No significant advantage was found over circular, symmetric
models.  The residuals similar to those shown in Fig.~4 did 
not generally decrease, unless the model was extended to
include asymmetric terms in addition.  

Asymmetric profiles were tested extensively on the supersampled 
PSF data of the selected frames used previously. In particular,
model 12 was used to probe parameter space and look for 
dependencies.
Smaller residuals than with any symmetric profile were found,
however different parameter values are needed for stellar
images in different locations on the detector as well as
for different CCD frames.

Figures 6 and 7 show some examples obtained with test runs
using models 9 and 11, respectively.  The amplitude, $c_{x}$ (Eq.~6),
of the asymmetric term along the $x$-coordinate (right ascension)
is approximated by a function linear with air temperature
and $x$ itself.  Image profiles are symmetric at low $x$ and
the largest asymmetry is seen at large $x$.
Similarly the amplitude of the asymmetry along the $y$-coordinate 
was estimated as a linear function of temperature and $y$.
Model 13 implements these preset values for $c_{x}$, and $c_{y}$, 
leaving only 6 free parameters, including the
$a, b$ profile widths along $x$ and $y$, respectively (elliptical
Lorentz base model).

\subsection{Double star fits}

Double star models solve simultaneously for at least 3 more 
parameters: the center coordinates, $x,y$ and amplitude, $A$ of 
the secondary component.
Again, the goal is to minimize the number of free parameters
as much as possible.  Thus, for example, a single parameter
for the background level is used.  Some double star models
also assume equal widths of the profiles of both components.
Table 1 gives more details (models 20 to 23). 

Critical for handling of blended images is the identification
of such cases and the determination of sufficiently
accurate starting parameters for the iterative, non-linear double 
star profile fit routine.  This process can easily ``go astray" 
due to the relatively large number of parameters and the small
number of pixels available with the critically sampled UCAC data.

The adopted criterion for detecting an object on a dark and flat corrected 
CCD frame is to have at least 2 connected pixels above a specified S/N
threshold level of $3 \sigma$ above mean background.  
For each such detected object a centroid position 
(center-of-light, 1st moments) is calculated as well as the 2nd moments.
The orientation of the major axis and image elongation, defined as
the ratio of major to minor axis are derived from these moments.  
An elongation of 1.0 means
a circular, symmetric image, otherwise the elongation is greater than 1.
An image profile fit with model 1 (circular, symmetric Gaussian) is 
performed on all objects to identify ``good" stars, and the mean 
image elongation of that CCD frame is calculated from the 2nd moment 
results of the ``good" stars only.
Images are typically slightly elongated due to guiding errors and
the CTE effect.

The double star routines are triggered if an objects elongation 
exceeds an adopted threshold of 12\% over the mean image elongation 
(for that CCD frame), and has a sufficient number of pixels ($\ge$10)
above the detection threshold level.
This elongation threshold was adopted as best compromise between
excluding false positives of single stars due to statistics 
and including as many as possible real double stars.
With some interpolation, pixel values are compared which lie on a
line through the center of light along the major axis as determined
earlier.  A search is made for 2 peaks along this line and
starting parameters (location and amplitudes) of the 2 components
are derived.  Starting parameters for the image profile width and
background value are taken from the overall CCD frame mean values.
If no 2 separate peaks can be detected, estimates for a possible
nearby, blended, secondary component are made based on the image 
profile width and brightness of the object under investigation.

A least-squares fit is attempted with these starting parameters
using model 23 (see Table 1), based on the Lorentz profile. 
The object is output as 2 components if reasonable starting parameters
could be derived, even if the double star fit failed.
A double star flag is assigned specifying the status of each successful
detection and/or fit of 2 components. 

A sample of newly detected UCAC3 doubles was observed with the
26-in speckle program \cite{26in} and a paper addressing accuracy
and reliability of UCAC3 double star data will be presented in
a separate paper \citep{u3ds}.

\section{UCAC3 PIXEL REDUCTION RUN}

\subsection{Algorithm}

The final reduction pipeline to process the UCAC3 pixel data
handles a specified range of CCD frames with a single selected 
master flat and pairs of low/high ADU master dark frames for each
standard exposure time (see above).
The input list of frames is sorted by exposure time, and frames
are then processed in that order, performing the following steps:

\begin{enumerate}
\item Read original, compressed pixel data file, determine mean 
     left/right background counts and flag saturated pixels.
\item Interpolate dark frame, apply dark and flat corrections,
        output processed image (round to 2-byte integers).
\item Flag pixels from bad pixel maps, detect and flag possible streaks
    (from shutter failure and bleeding images).
\item Pass 1:
  \begin{enumerate}
  \item Detect objects ($3 \sigma$ above mean background for at
    least 2 connected pixels).
  \item  Classify objects including 1st and 2nd moments, identify possible 
   blended images (doubles).
  \item Perform image center fits on all objects with model 1 (Gaussian).
  \end{enumerate}
\item Intermission 1: 
    \begin{enumerate}
    \item Identify ``good" stars over entire CCD frame, based on model 1
      fit results.
    \item  Derive mean image profile width, mean image elongation,
       $\alpha, \beta$ profile shape parameters, and 
       radii for aperture photometry.
    \end{enumerate}
\item Pass 2:
  \begin{enumerate}
  \item Perform circular, symmetric Lorentz profile fit (model 5).
  \item Calculate double star starting parameters and perform fit (model 23).
  \item Perform asymmetric profile fit (model 13).
  \end{enumerate}
\item Intermission 2:
  \begin{enumerate}
  \item Select ``good" stars from model 13 fit results.
  \item Derive mean width of profiles ($a, b$) of elliptical part of model 13
      and fixed parameters for model 14.
  \end{enumerate}
\item Pass 3: Perform model 14 fit with further parameter restrictions.
\item Perform aperture photometry.
\item Derive model magnitudes from each successful model fit.
\item Output all fit results for each frame to a separate file.
\end{enumerate}

The profile fits are performed with pixels inside a circular aperture
centered on the best known position at the time.  The radius of this
aperture was adopted to be twice as large as the radius of the area
of pixels above the threshold from the image detection step.
For all astrometric image profile fits the local background parameter
is a free fit parameter for each single star or binary pair.

Note, the astrometric fit based on the Lorentz profile is performed with 
5 free parameters, the same number of parameters as used for a Gaussian 
profile model.
The major difference is that here a model profile is selected that does 
better match the data with the profile shape being slightly different for 
CCD frames taken under different seeing conditions (average image width).
The first step in this reduction process (finding $\alpha, \beta$) merely
is a quantitative way to make a ``good guess" about which model profile 
to use.

For the aperture photometry all pixels within 4 times the mean radius of 
the image profiles (Gaussian fit of ``good" stars) of that CCD frame were 
used to determine the flux of a target.  An annulus with 12 and 16 times
this mean profile radius served for the background determination.
The background value is determined from the peak of the histogram of 
background pixels (weighted mean of bins which exceed 50\% of the
smoothed histogram peak value).

\subsection{Processing and Results}

The re-processing of the pixel data involved 5 different Linux 
workstations (mostly single processor at about 2 GHz),
which most of the time ran in parallel for about a month 
working on a section of the UCAC frames each.  
Individual binary files contain the output data for each CCD frame.
The number of objects per frame ranged from 37 to 75,549
with a median of 1397.
A data record of 136 bytes contains the results for each detected object,
including selected items from the moment analysis, the parameters
of the 4 model fits, their errors, and flags.  Data items were converted
to integers of 1, 2, or 4 byte lengths with appropriate scaling.
A total of over 4 TB of compressed pixel data went into this process,
producing a total of 80 GB of binary $x,y$ data, the results of 271,428
CCD exposures.
These files were later extended by 8 bytes per record to arrive at 
the final $x,y$-data output files.  The additional data contain 
information about nearest neighbors, and identification of possible 
doubles that are not blended.

\subsection{Analysis of the Results}

How good are the resulting $x,y$ data?
An example of the internal fit precision is presented in Figure 8.
The formal standard error of the $x$-center coordinate is shown
as a function of instrumental magnitude.  Data for the same CCD
frame are shown for 3 different image profile fit models (1, 5, and 14).
The results for model 13 look very similar to those of model 14. 
The results for the $y$ coordinate are similar to those of the 
$x$-coordinate.  Saturation occurs at about magnitude 8.
The unit is milli pixel (mpx), with 1 mpx = 0.9 mas.
For the unsaturated, high S/N stars (8 to 10 mag) per coordinate 
precisions of 
about 6 mpx, 4 mpx, and 3 mpx are reached for fit models 1, 5, and 14,
respectively, from a single CCD frame observation of good quality.

The repeatability of observations was tested frequently with
the same field in the sky observed twice within minutes and
the telescope being on the same side of the pier.
A weighted, linear transformation between the sets of $x,y$ data of
such a pair of 100 s exposures was performed and the scatter in
the $x$ coordinate of stars plotted as a function of instrumental
magnitude in Figure 9.
Again, results from different profile fit models (pfm) are shown
as indicated.  This scatter includes the errors from both CCD frames.
Assuming equal error contribution, the repeatability error of these
observations is thus about 15 mpx / $\sqrt{2}$ $\approx$ 11 mpx 
$\approx$ 10 mas per coordinate and single observation for well
exposed stars, almost independent of the profile fit model.
These results are consistent with the first observations at this
telescope using a CCD camera \citep{aqUCA}.
The observed error is significantly larger than the internal fit 
precision (of bright stars) due to atmospheric turbulence.
A scatter that is about a factor of 2 larger is observed
in similar CCD frame pairs of 25 s exposure time as compared 
to the 100 s frames.

The flip observations, with the telescope on one side of the pier
then on the other, provide pairs of frames that are rotated by 
$180^{\circ}$ with respect to each other.  
A linear transformation between the 2 sets of $x,y$ data of each 
frame pair taken on the same field in the sky gives residuals revealing
systematic errors as a function of magnitude or coma-like terms 
(product of magnitude and positional coordinates).  
An example is shown in Figure 10 for such a pair of frames
with 100 s exposure, before applying corrections.
Results are derived from the same pixel data but 
using all 4 different profile fit models as indicated.
Each data point is the mean for 16 stars.
The model 5 and 13 results show a somewhat tighter distribution than
those of models 1 and 14.
Unfortunately a similar amplitude of the systematic errors
is present in the data from all 4 models, while the hope has been 
that the asymmetric profile fits would have mitigated this problem.
Even larger systematic errors (about 200 mas) are seen in short
exposure frames.
These and other systematic errors will be investigated in great
detail with the help of reference stars, as described in another
paper of this series \citep{ared}.
Empirical corrections will be derived for these and other systematic
errors in the UCAC data at that time.  These corrections effectively 
reduce these types of systematic errors by about a factor of 10 
(as compared to what is seen in Fig.~10) for the published catalog 
star positions.

\section{DISCUSSION AND CONCLUSIONS}

The detection threshold for blended, double star images is affected by
the gradient of inherent image elongation along the $x$ axis due to
the poor CTE.  Similarly fit results of double stars will have a slight
bias depending on the $x$ pixel coordinate.  Almost symmetric images
are seen on the left side of a CCD frame, while CTE elongated, asymmetric
images are seen on the right side.  In all cases the same, symmetric double
star profile is used for a fit.  In addition, a slight image elongation
is typically added from guiding effects.  Nevertheless, an important
first step for detecting double stars and deriving useful parameters
has been accomplished for UCAC3.  Results from external comparisons 
will be presented in \citep{u3d}. 

The additional first order parameters to describe image
asymmetry (relative amplitude of a term linear with pixel coordinate)
could not be correlated well to any other parameters,
contrary to the $\alpha$ and $\beta$ shape parameters used
in the Lorentz profile model.
Even such a complex model applied without approximations
of the asymmetric parameters (i.e.~use of many free fit parameters)
leaves significantly large residuals (see Fig.~4 bottom) in
the supersampled, stacked PSFs.
Applying a model with 7 or more parameters to the (not supersampled)
original pixel data for each star is not an option with the critical 
sampling of the UCAC data (too few pixels per image).

This situation calls for a purely empirical PSF model by using the
supersampled, stacked, observed profiles to generate a template.
Using purely empirical PSFs as templates to fit observed stellar
images in UCAC data, however, was not considered a viable option.
Many UCAC frames have a low number of stars, in particular the
number of stars with high S/N ratio needed for this
approach is very small (a few) in many areas of the sky.
Furthermore, the PSF changes significantly over the area of the
detector (mainly along the $x$ coordinate) due to the poor CTE
and, is a function of the profile width (FWHM), the guiding of
each exposure, the CCD temperature and probably other factors.
Splitting up the data into so many categories was not an option.
In general a purely empirical PSF will be asymmetric to some
degree, also caused by imperfect guiding.
Such an approach would mean a different definition of an image
center for different CCD frames with additional variations
over the field of view, not desirable for astrometry.
Thus the use of models 13 and 14 is a compromise driven by the need
for a more complex model without having a large enough sampling to
support it.

The very high precision of the well exposed stellar images
seen in the internal fit errors per coordinate unfortunately
does not translate into similarly small external errors.
Internal errors of under 5 mas per coordinate and exposure are
seen, however the repeatability of such observations is already
degraded to about 10 mas and more due to the atmosphere for our
long exposures of 100 to 150 s.  For the short exposures
the positional errors are further increased by about a factor of 2.
However, the UCAC data are still limited by remaining systematic
errors, resulting in an error floor of just under 20 mas per coordinate
for the mean CCD observations (4 images) for stars in the R = 10
to 14 mag range, as will be shown in the astrometric
reduction paper \citep{ared}.

The symmetric profile model 5 $x,y$ center coordinates
have been adopted as the baseline for these astrometric reductions.
It could be argued that the derived $\alpha$ and $\beta$ shape
parameters used in model 5 are not accurately enough known, 
based on the approximation as described above.
However, very good astrometric results have already been obtained
with the Gaussian model (UCAC2, and many other traditional,
astrometric catalog projects).  The Lorentz profile adopted here,
with a somewhat imperfect representation of shape parameters,
is a far better representation of the observed image profile 
than is the Gaussian model.
Both are symmetric profile functions with the same number of
free parameters, so no detrimental effect is expected when
choosing model 5 (Lorentz profile) over model 1 (Gauss profile).

Effects from image asymmetry will be investigated and corrected
by analyzing the residuals with respect to reference stars.
No significant benefit for the overall astrometric accuracy
has been found so far by using the asymmetric profile models 
investigated here, particularly with respect to solving magnitude 
and coma-like terms.
For the baseline UCAC reductions a symmetric PSF model is applied 
to asymmetric image profiles with subsequent systematic error 
corrections of the celestial coordinates using reference stars.

\acknowledgments

The entire UCAC team is thanked for making this all-sky survey a reality.
In particular mention should be made of,
Charlie Finch for discussions and assistance with data
structure implementations and
Gary Wycoff for preparatory work including frame selection tasks.
National Optical Astronomy Observatories (NOAO) is acknowledged for
IRAF, Smithonian Astrophysical Observatory for DS9 image display software,
and the California Institute of Technology for the {\em pgplot} software.
More information about this project is available at \\
\url{http://www.usno.navy.mil/usno/astrometry/}.





\clearpage



\begin{figure}
\epsscale{1.00}
\plotone{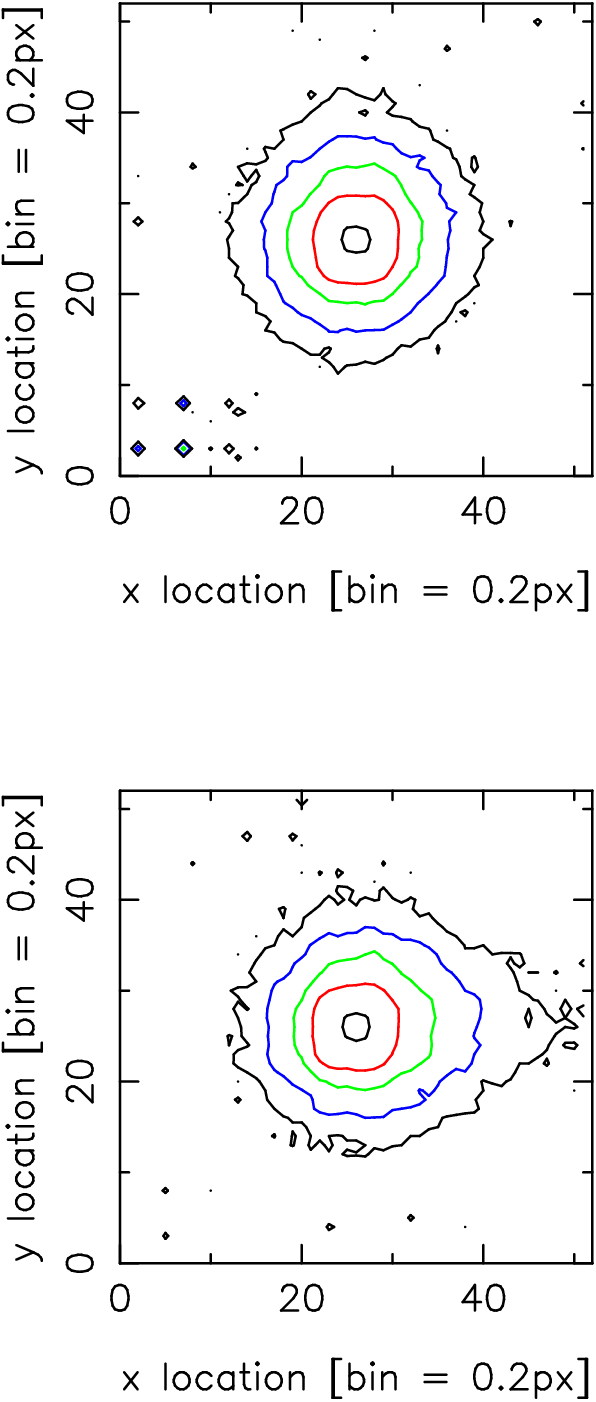}
\caption{Contour plot of supersampled (see text) images
  of CCD frame 134473.  Average data from near the output register 
  (top) and farther away from it (bottom) are shown.  This illustrates
  the low charge transfer efficiency problem of the detector leading to
  images which are asymmetric as a function of the $x$ pixel coordinate.
  The contour levels are at 90\%, 30\%, 10\%, 3\% and 1\% of the peak
  intensity.}
\end{figure}

\begin{figure}
\epsscale{1.00}
\plotone{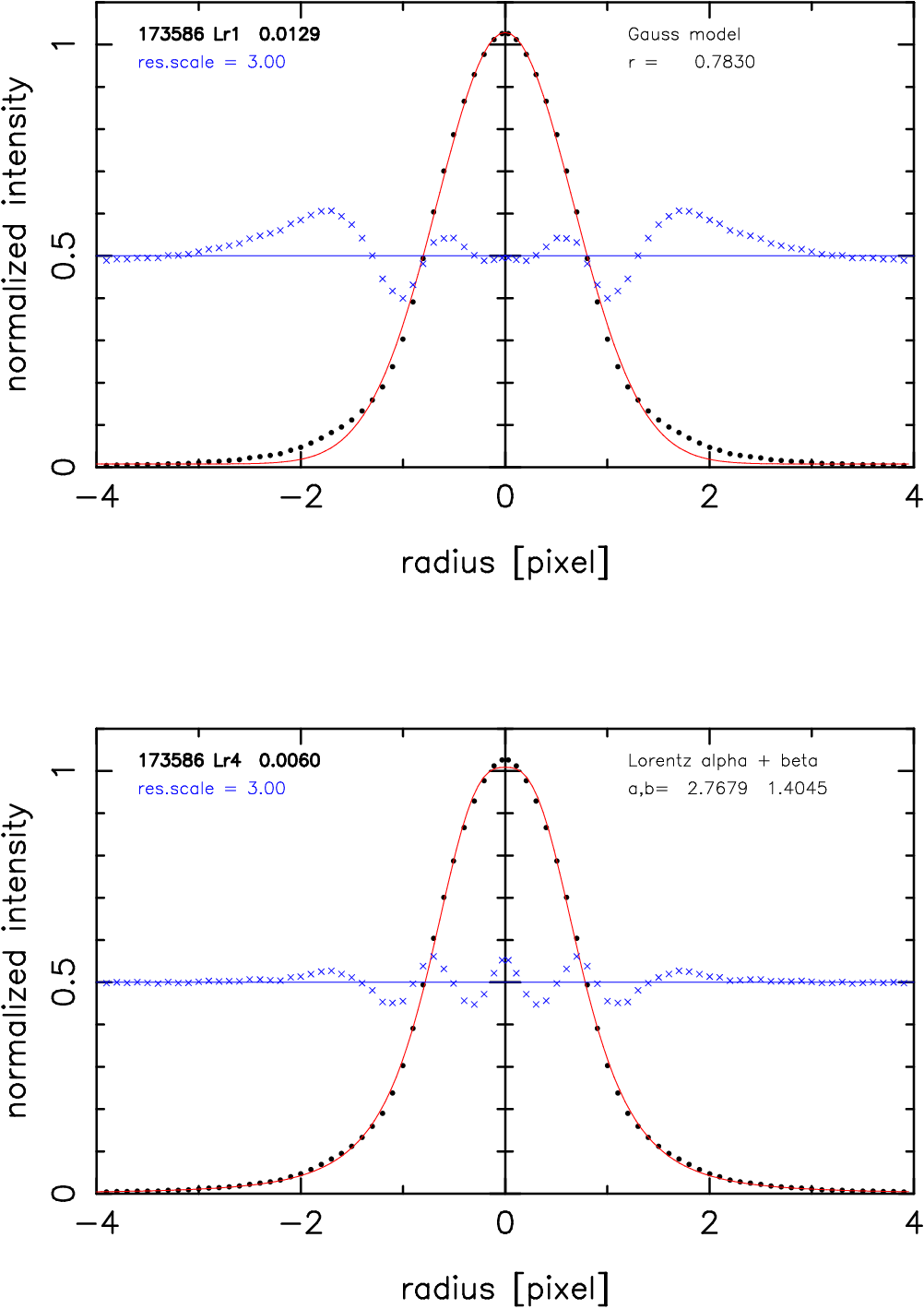}
\caption{Radial profile plot of supersampled (see text) images
  on the left side of CCD frame 173586.  The same data points 
  are shown in both diagrams with a fit of a Gaussian (top) and 
  Lorentz model function (bottom).
  Residuals are scaled with a factor of 3 and shown with
  an intensity offset of 0.5 for better readability.}
\end{figure}

\begin{figure}
\epsscale{1.00}
\plotone{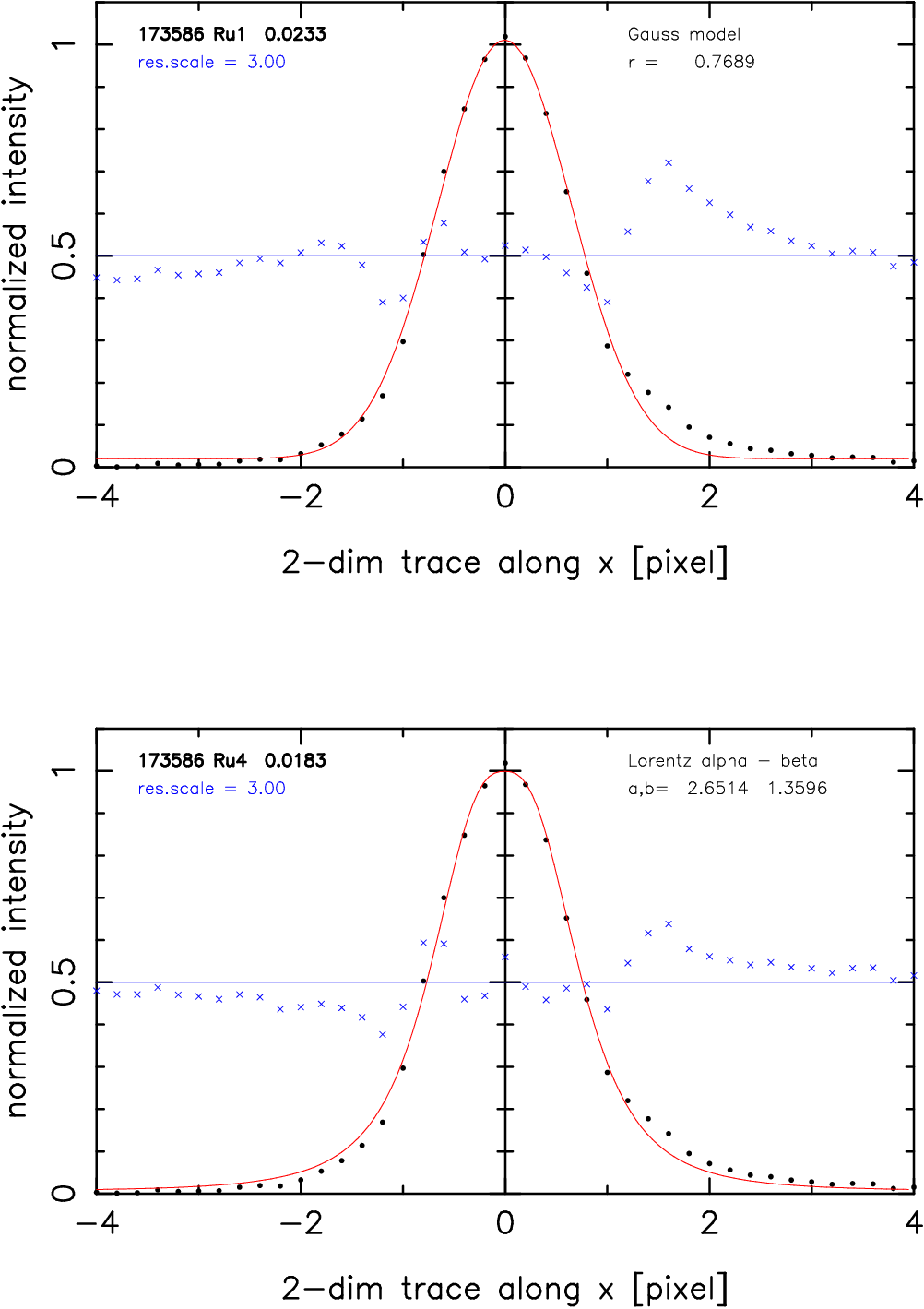}
\caption{Profile plot along $x$-axis of supersampled images
  on the right side of CCD frame 173586.  The same data points 
  are shown in both diagrams with a fit of a Gaussian (top) and 
  Lorentz model function (bottom).
  Residuals are scaled with a factor of 3 and shown with
  an intensity offset of 0.5 for better readability.}
\end{figure}

\begin{figure}
\epsscale{1.00}
\plotone{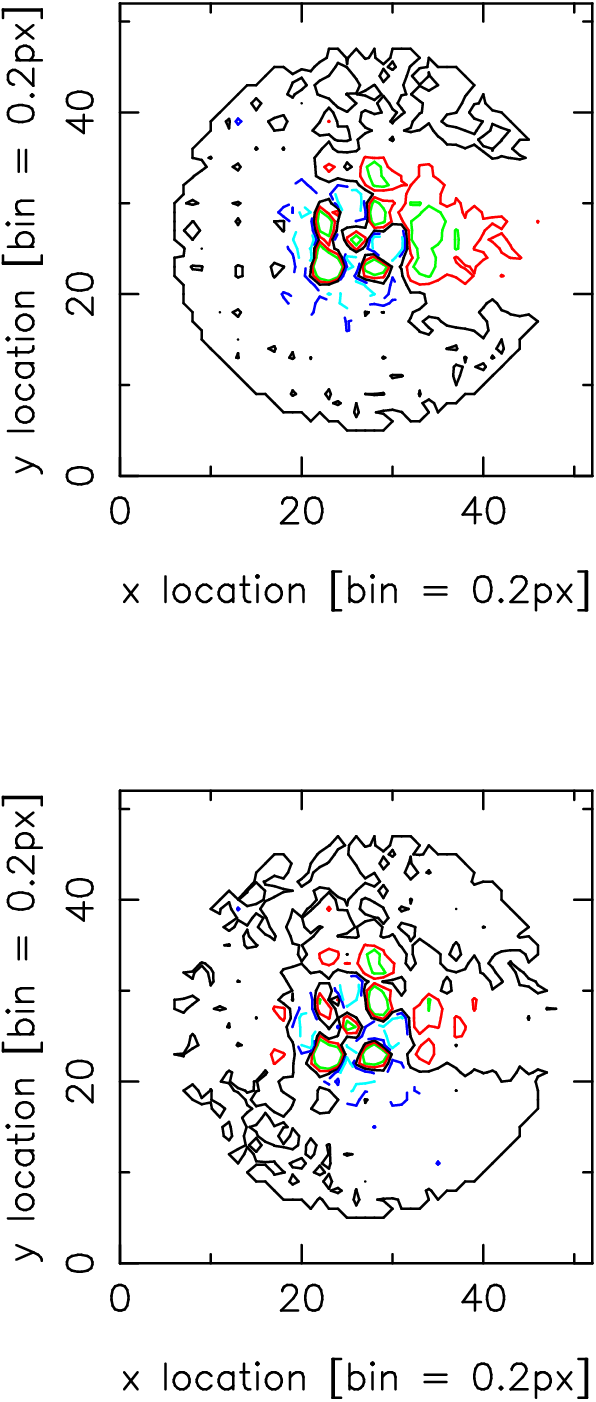}
\caption{Residual contour plot of a supersampled PSF from
  stellar images on the right side of CCD frame 134473. 
  The same data are used for both plots; model 4
  (symmetric Lorentz profile) fit (top), and model 9 
  (elliptic Lorentz profile plus asymmetry terms
  along $x$) (bottom) fits were applied, respectively.
  Contour levels are light blue $-3$\%, dark blue $-1.5$\%,
  black 0, red $+1.5$\%, and green $+3$\%.} 
\end{figure}

\begin{figure}
\epsscale{1.00}
\plotone{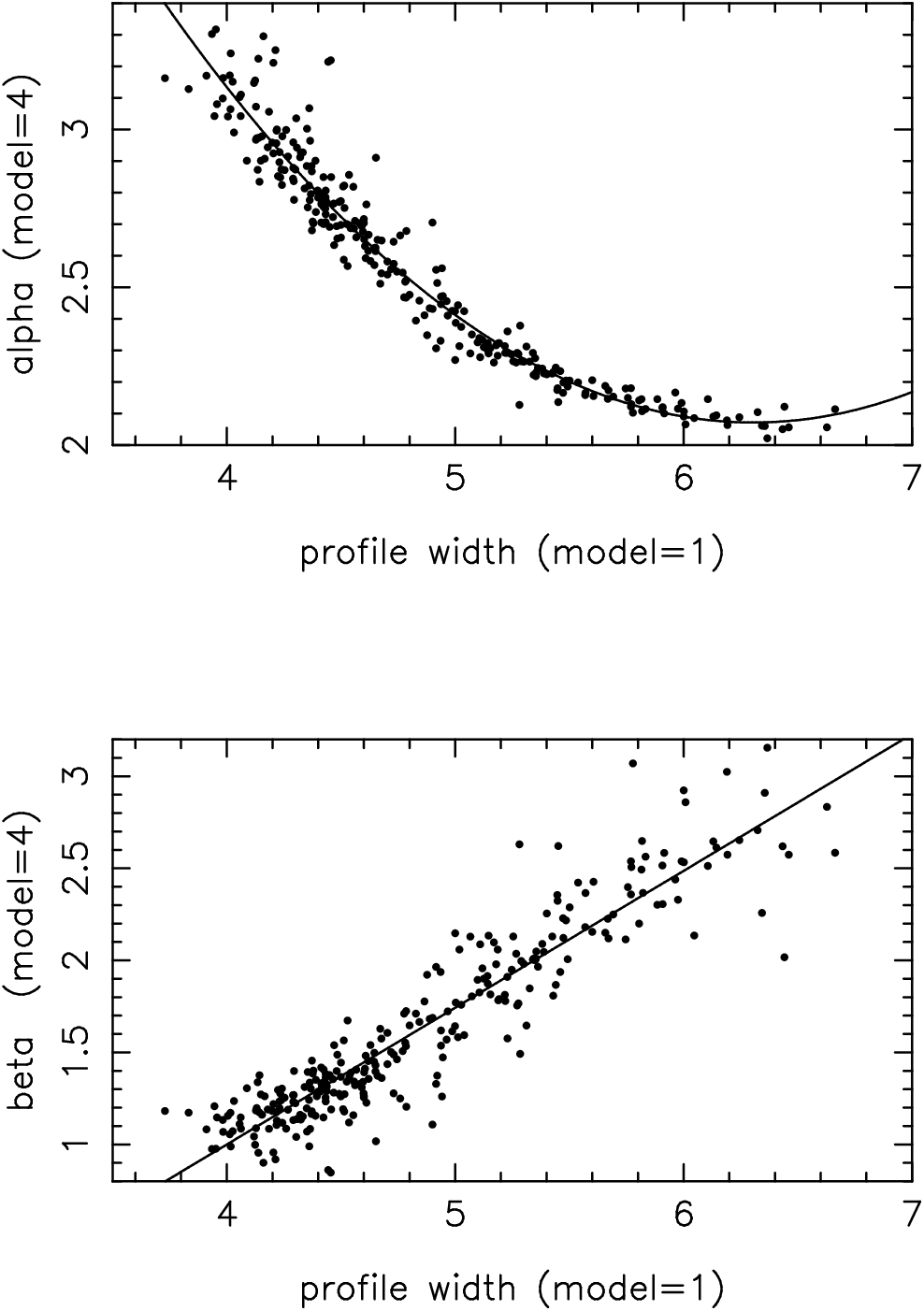}
\caption{Alpha (top) and beta (bottom) image profile shape 
  parameters of the Lorentz model 4 as a function of 
  the image width (radius) as determined by model 1 (Gauss)
  for the supersampled data (1 bin = 0.2 pixel).
  The alpha parameter is fit by a 2nd order polynomial, while
  the beta parameter is well represented by a linear dependency.}
\end{figure}

\begin{figure}
\epsscale{1.00}
\plotone{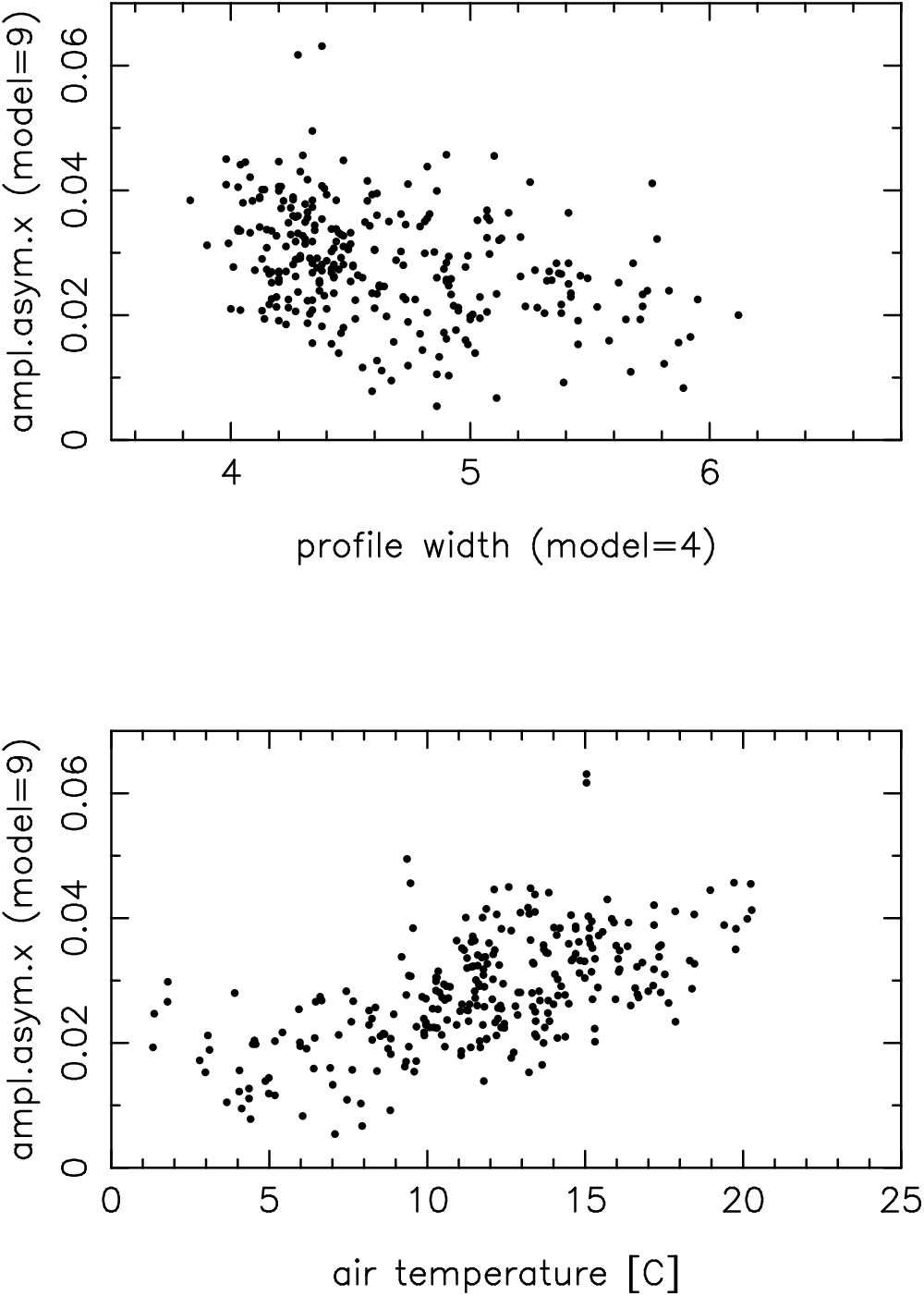}
\caption{Examples of the asymmetry along 
  $x$ parameter dependencies from supersampled data.
  The amplitude of the asymmetry along $x$ (RA) as a function of 
  the air temperature (bottom) and as a 
  function of the image profile radius (top)
  are among the strongest correlations found.}
\end{figure}

\begin{figure}
\epsscale{1.00}
\plotone{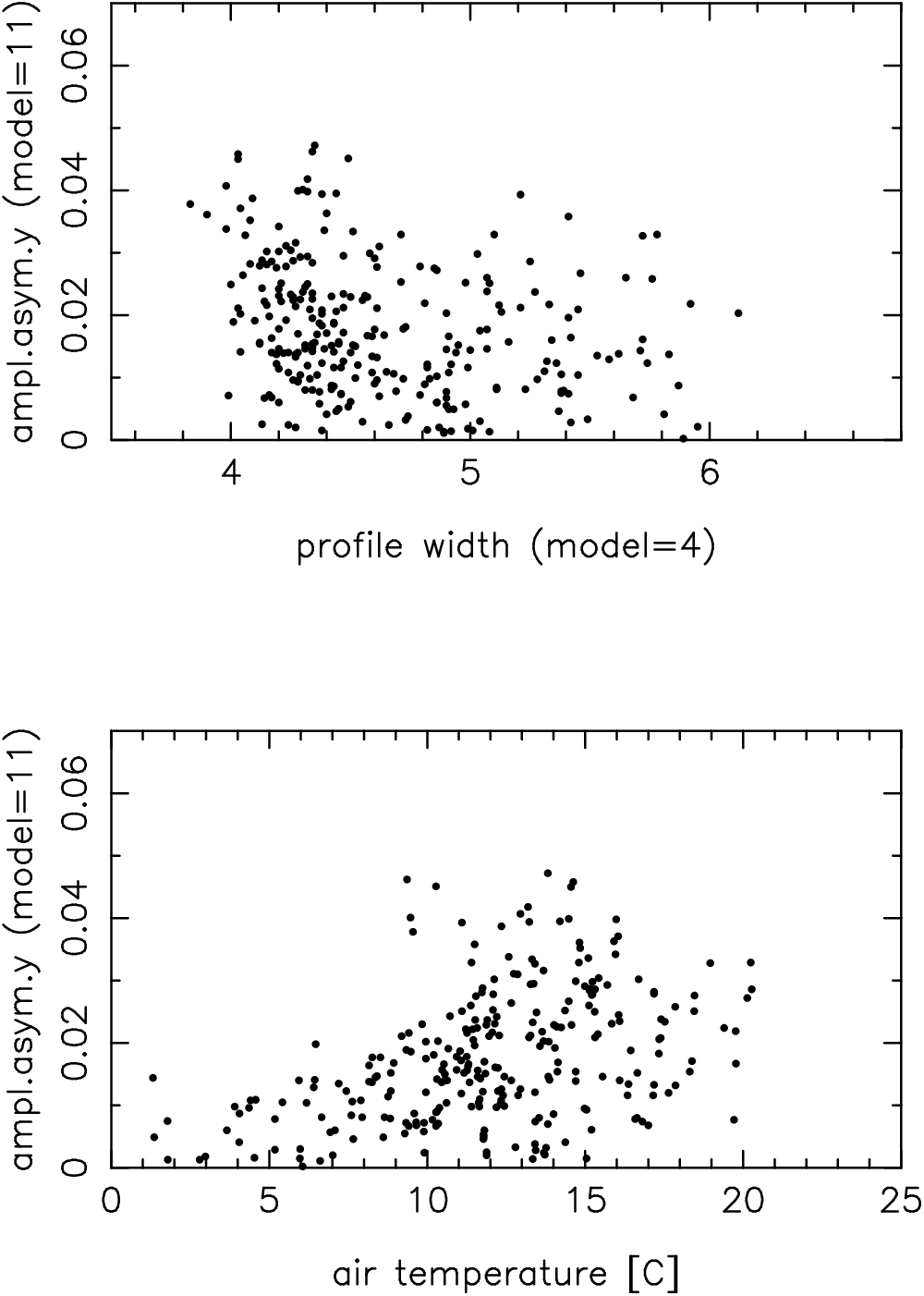}
\caption{Similarly to the previous figure, the dependency of
  the amplitude of image asymmetry along $y$ (Dec) is shown here.}
\end{figure}

\begin{figure}
\epsscale{1.00}
\plotone{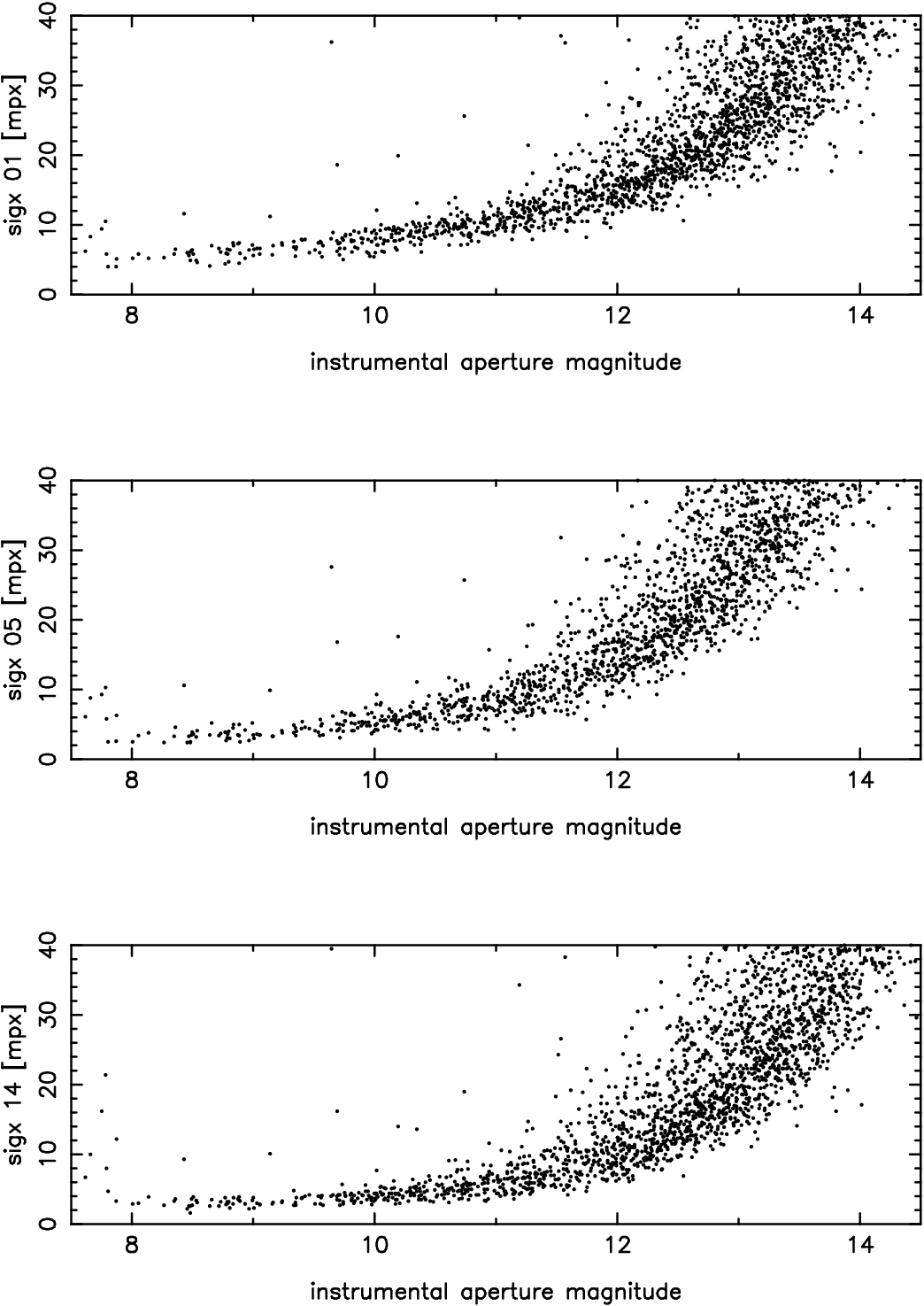}
\caption{Internal fit precision (milli pixel = 0.9 mas) along 
  the $x$ coordinate as a function of aperture magnitude of a 
  long exposure sample frame 53554  with over 4000 stars.
  The plots zoom in on the high
  precision, high S/N area (bright stars).  Results for different 
  fit models are shown: model 1 (Gaussian) on top, model 5 (Lorentz)
  in the middle, and model 14 (asymmetric profile) on the bottom.}
\end{figure}

\begin{figure}
\epsscale{1.00}
\plotone{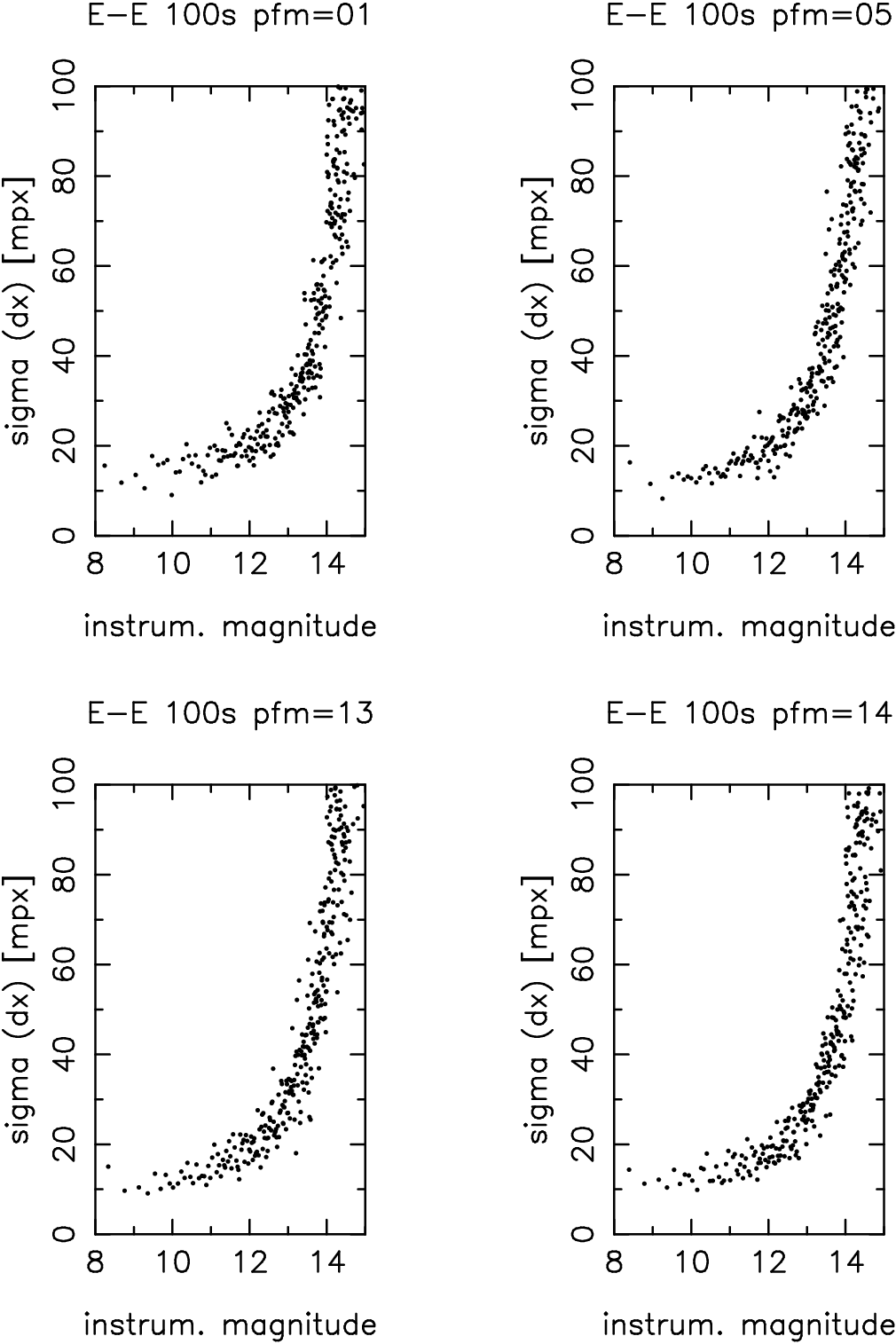}
\caption{Standard error (sigma) of the $x$ coordinate as 
  a function of instrumental model magnitude from the
  comparison of 2 100-second exposures taken of the same
  field in the sky within minutes.  
  Results are shown from the same CCD frame pair but for
  different image profile fit models (pfm) as indicated.
  The error shown is the combined error of both
  exposures, dominated by atmospheric effects.
  Each dot represents the RMS average of 16 stars.
  Unit is milli pixel (1 mpx = 0.9 mas).}
\end{figure}

\begin{figure}
\epsscale{1.00}
\plotone{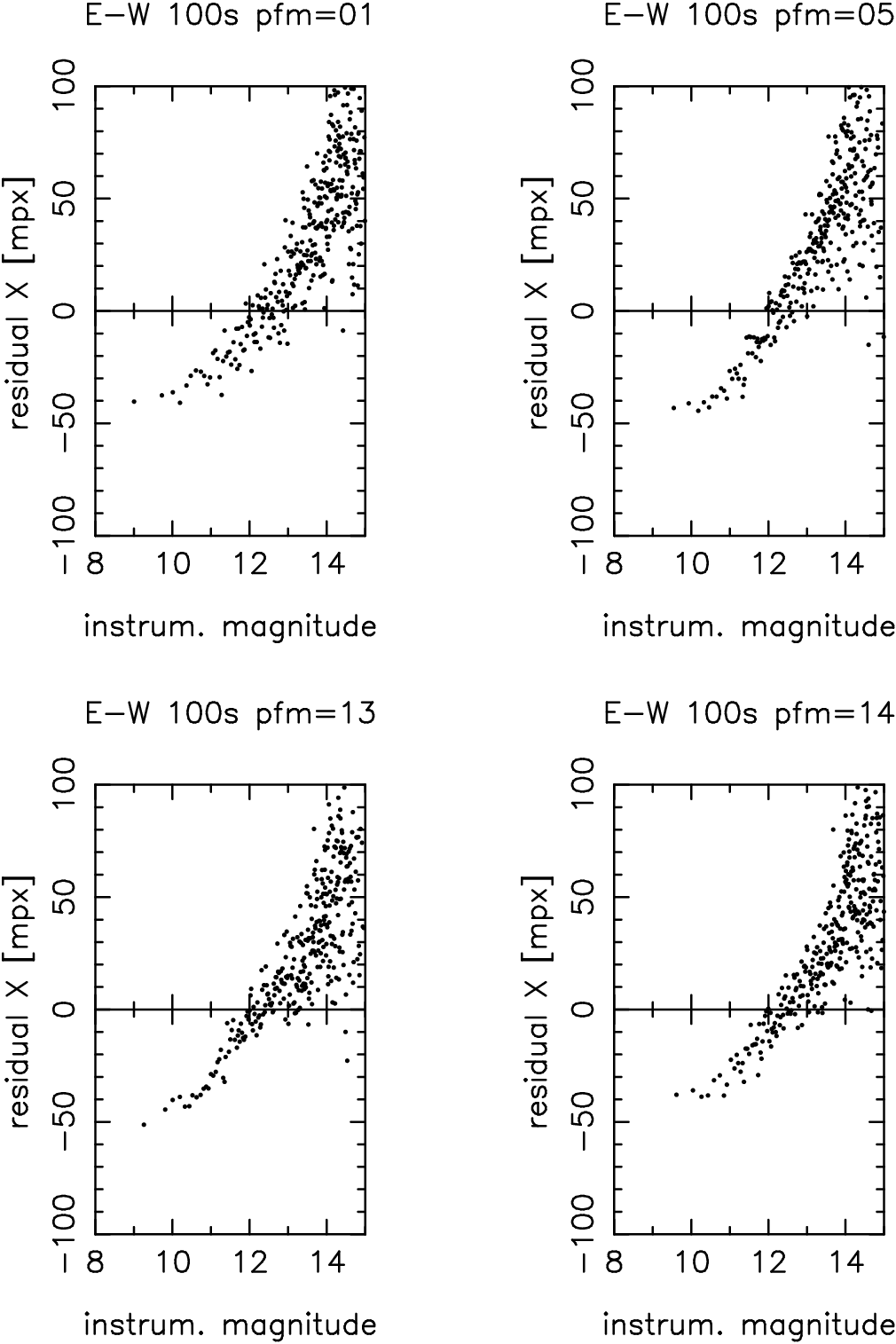}
\caption{Systematic differences between the $x$ coordinates 
  of a pair of 100-second exposures taken from opposite sides
  of the telescope pier (rotated by $180^{\circ}$ with
  respect to each other) as
  a function of instrumental model magnitude, after a
  linear transformation of the $x,y$ data but without any
  other corrections applied.
  Results are shown from the same CCD frame pair but for
  different image profile fit models (pfm) as indicated.
  Each dot represents the mean over 16 stars.
  Unit is milli pixel (1 mpx = 0.9 mas).}
\end{figure}








\clearpage

\begin{table}
\begin{center}
\caption{Description of image profile models used in UCAC3 reductions and
         tests.}
\vspace*{5mm}
\begin{tabular}{lcccccccccccccccccc}  
\tableline\tableline
           & \multicolumn{18}{c}{profile fit model number} \\
                            & 1 & 2 & 3 & 4 & 5 & 6 & 7 & 8 & 9 &10 &11 &12 &13 &14 &20 &21 &22 &23\\
\tableline
base model\tablenotemark{a} & G & D & L & L & L & G & G & L & L & L & L & L & L & L &Gd &Gd &Ld &Ld\\
symmetry\tablenotemark{b}   & c & c & c & c & c & e & e & e &e,a&e,a&e,a&e,a&e,a&e,a& c & c & c & c\\
total number of parameters  & 5 & 6 & 7 & 7 & 7 & 6 & 7 & 8 & 9 & 9 & 9 &10 &10 &10 & 8 & 9 &11 &11\\
numb.of free fit parameters & 5 & 6 & 6 & 7 & 5 & 6 & 7 & 8 & 9 & 7 & 9 & 8 & 6 & 4 & 8 & 9 & 9 & 8\\
\tableline
$x,y$ center, ampl., backgr.\tablenotemark{c}&f&f&f&f&f&f&f&f&f & f & f & f & f & f & f & f & f & f\\
profile width\tablenotemark{d}&f& f & f & f & f & f & f & f & f & f & f & f & f & p & f & f & f & f\\
elliptical axis orientation &   &   &   &   &   &   & f &   &   &   &   &   &   &   &   &   &   &  \\
1st shape parameter         &   & f & f & f & p &   &   & f & f & p & f & p & p & p &   &   & p & p\\
2st shape parameter         &   &   & p & f & p &   &   & f & f & p & f & p & p & p &   &   & p & p\\
asymmetry $x$ amplitude     &   &   &   &   &   &   &   &   & f & f &   & f & p & p &   &   &   &  \\
asymmetry $y$ amplitude     &   &   &   &   &   &   &   &   &   &   & f & f & p & p &   &   &   &  \\
$x,y$, amplitude secondary  &   &   &   &   &   &   &   &   &   &   &   &   &   &   & f & f & f & f\\
profile width secondary     &   &   &   &   &   &   &   &   &   &   &   &   &   &   &   & f & f &  \\
\tableline
\end{tabular}
\tablenotetext{a}{G = Gaussian, D = double exponential, L = Lorentz, d = double star}
\tablenotetext{b}{c = circular symmetric, e = elliptical, 
                  a = incl.~asymmetry}
\tablenotetext{c}{f = free fit parameter, p = preset}
\tablenotetext{d}{1 parameter for circular symmetric, 2 for elliptical}
\end{center}
\end{table}



\end{document}